\pdfoutput=1
\RequirePackage{ifpdf}
\ifpdf % We~are running pdfTeX in pdf mode
\documentclass[pdftex]{sigma}
\else
\documentclass{sigma}
\fi

\numberwithin{equation}{section}

\newtheorem{Theorem}{Theorem}[section]
\newtheorem*{Theorem*}{Theorem}

\newtheorem{Proposition}[Theorem]{Proposition}
 { \theoremstyle{definition}

 }

\def\ben{ \begin{eqnarray}}
\def\en{ \end{eqnarray}}

\begin{document}
\allowdisplaybreaks

\newcommand{\arXivNumber}{2201.09576}

\renewcommand{\PaperNumber}{094}

\FirstPageHeading

\ShortArticleName{Equivalent Integrable Metrics on the Sphere with Quartic Invariants}

\ArticleName{Equivalent Integrable Metrics on the Sphere\\ with Quartic Invariants}

\Author{Andrey V.~TSIGANOV}

\AuthorNameForHeading{A.V.~Tsiganov}

\Address{St.~Petersburg State University, St.~Petersburg, Russia}
\Email{\href{mailto:email@address}{andrey.tsiganov@gmail.com}}

\ArticleDates{Received March 31, 2022, in final form December 04, 2022; Published online December 06, 2022}

\Abstract{We discuss canonical transformations relating well-known geodesic flows on the cotangent bundle of the sphere with a set of geodesic flows with quartic invariants. By adding various potentials to the corresponding geodesic Hamiltonians, we can construct new integrable systems on the sphere with quartic invariants.}


\Classification{37J35; 70H06; 70H45}

\section{Introduction}

In the study of metric spaces, there are various notions of two metrics on the same underlying space $Q$ being ``the same'', or equivalent. For instance, there are topologically equivalent metrics and strong equivalent metrics~\cite{bish80}. In the Riemannian geometry two metrics are projectively equivalent if their geodesics coincide, in K\"ahler geometry two metrics are c-projectively equivalent if their $J$-planar curves coincide and so on, see \cite{kiy11,mat20,tab84} and references within.

In symplectic geometry, it is natural to say that two metrics $\mathrm g$ and $\mathrm g'$ on the configuration space $Q$ are equivalent if the corresponding geodesic Hamiltonians
\begin{equation} \label{gHam}
T=\sum_{i,j=1}^n \mathrm g_{ij}(q) p_ip_j\qquad \mbox{and}\qquad
T'=\sum_{i,j=1}^n \mathrm g'_{ij}(q) p_ip_j
\end{equation}
are related by some transformation of the phase space
\begin{equation}\label{g-trans}
\rho\colon \  T^*Q\to T^*Q
\end{equation}
preserving canonical symplectic form $\omega={\rm d}p\wedge {\rm d}q$. Here $q=q_1,\dots,q_n$ are coordinates on $Q$ and $p=p_1,\dots,p_n$ are
are fibrewise coordinates with respect to the cotangent vectors ${\rm d}q_1,\dots,{\rm d}q_n$.
Well-known examples of such canonical transformations are point transformations and non-point transformations in $T^*{\mathbb R}^n$
\[
\rho\colon \ q_i\to p_i\qquad\mbox{and}\qquad p_i\to -q_i ,\qquad i=1,\dots,n ,
\]
relating two geodesic Hamiltonians (\ref{gHam}) when both metrics are the homogeneous polynomials of second order in coordinates.
We aim to construct and classify other non-point canonical transformations relating to two polynomials of the second order in momenta~(\ref{gHam}).

Canonical transformation preserves the form of canonical Poisson brackets, which allows us to obtain new integrable geodesic flows by using the following algorithm:
\begin{itemize}\itemsep=0pt
 \item take some known integrable geodesic flow with Hamiltonian $T=T_1$ and independent integrals of motion $T_2,\dots,T_n$ in the involution
 \[\{T_i,T_j\}=0 ,\qquad i,j=1,\dots,n;\]
 \item take non-point canonical transformation $\rho$ (\ref{g-trans}), which maps geodesic Hamiltonian $T$ to geodesic Hamiltonian $T'$, and calculate a set of independent functions $\rho(T_k)$ in the involution on $T^*Q$ with respect to the same canonical Poisson brackets
 \[\{\rho(T_i),\rho(T_j)\}=0 ,\qquad i,j=1,\dots,n ;\]
 \item compute $n-1$ functions $K_m$ on integrals of motion $\rho(T_k)$, so that functions $K_m$ are polynomials in momenta, which simplifies all further calculations;
 \item find potential $V(q)$ solving equations
 \[\{H_i,H_j\}=0 ,\qquad i,j=1,\dots,n,\]
 where $H_1=T'+V(q)$ and $H_m=K_m+W_m(p,q)$, with respect to $V$ and polynomials in momenta $W_m$;
 \item calculate new integrable metric $\tilde{\mathrm g}$ on $Q$ by using Maupertuis principle
 \begin{equation}\label{mau-metr}
 \tilde{H}=\frac{T'}{h-V}=\sum_{i,j=1}^n \tilde{\mathrm g}_{ij}(q) p_ip_j .
 \end{equation}
\end{itemize}
The main unsolved problems in this method are the construction of the non-point canonical transformations $\rho$ (\ref{g-trans}) relating a given quadratic polynomial~$T$ with other quadratic polynomial~$T'$ and computation of the applicable to the Maupertuis principle polynomials in momenta~$K_m$.
Several canonical transformations $\rho$ were obtained in the framework of algebraic geometry for the 2D Euclidean space in \cite{ts15b,ts17,ts18d,ts18g}, for the 2D sphere in \cite{ts15,ts15a,ts17v,ts17c} and for the 2D ellipsoid in~\cite{ts17ell}.

In this note, we present canonical transformation $\rho$ (\ref{g-trans}) on the cotangent bundle to $(n-1)$-dimensional sphere $S^{(n-1)}$ using globally defined coordinates on the ambient space~${\mathbb R}^n$. At $n=3$ this transformation was obtained in \cite{ts17v,ts17c} in terms of the locally defined coordinates on the sphere.
Because we only want to prove the existence of such non-point canonical transformations~$\rho$ (\ref{g-trans}) and their applicability to the construction of new integrable metrics and so-called magnetic Hamiltonians $H_1=T'+V$ with generalized potential $V$ depending on velocities
\[V=\sum_{i=1}^n u_i(q)p_i+U(q) ,\]
we do not discuss the properties of obtained integrable systems, the curvature of the metrics, etc.

\section{Non-point canonical transformations}

 Let us consider Cartesian coordinates $x=(x_1,\dots, x_n)$ in Euclidean space $\mathbb R^{n}$ and the conjugated momenta $p_{x_i}$ on $T^*\mathbb R^{n}$, so that
\[
\{x_i,x_j\}'=\{p_{x_i},p_{x_j}\}'=0 ,\qquad \{x_i,p_{x_j}\}'=\delta_{ij} ,\qquad i,j=1,\dots,n.
\]
The unit $(n-1)$-dimensional sphere $\mathbb S^{(n-1)}\subset \mathbb R^n$ and its cotangent bundle $T^*\mathbb S^{(n-1)}\subset T^*\mathbb R^n$ are defined via constraints
\begin{equation} \label{f-con}
F_1=x_1^2+\cdots+x_n^2=1 , \qquad
F_2=x_1p_{x_1}+\cdots+x_np_{x_n}=0 .
\end{equation}
Induced symplectic structure on $T^*\mathbb S^{n-1}$ is given by the Dirac--Poisson bracket
\[
\{f,g\}=\{f,g\}'-\dfrac{\{F_1,f\}'\{F_2,g\}'-\{F_1,g\}'\{F_2,f\}'}{\{F_1,F_2\}'},
\]
which reads as
\begin{equation}\label{d-br}
\{x_i,x_j\}=0 ,\qquad \{x_i,p_{x_j}\}=\delta_{ij}-x_ix_j ,\qquad \{p_{x_i},p_{x_j}\}=x_jp_{x_i}-x_ip_{x_j} .
\end{equation}
Images of these variables $(x,p_x)$ we denote as $y=(y_1,y_2,y_3)$ and $p_y=(p_{y_1},p_{y_2},p_{y_3})$
\begin{Proposition}
Consider the following mapping of the cotangent bundle $T^*\mathbb S^{(n-1)}$
\[
\rho_b\colon \ (x_i,p_{x_i})\to (y_i,p_{y_i}) ,\qquad i=1,\dots,n,\]
defined by equations
\begin{equation}\label{g-trans2}
y_i= \sqrt{\frac{b_i}{\mathcal H}} p_{x_i}\qquad\text{and}\qquad x_ip_{x_i}+y_ip_{y_i}=0 ,
\end{equation}
where
\[\mathcal H=b_1p_{x_1}^2+\cdots +b_n p_{x_n}^2\qquad\text{and}  \qquad b_i>0 .\]

This mapping preserves constraints
\begin{gather*}
F_1=x_1^2+\dots+x_n^2=1=y_1^2+\cdots+y_n^2 ,\\
F_2=x_1p_{x_1}+\dots+x_np_{x_n}=0=y_1p_{y_1}+\dots+y_np_{y_n} ,
\end{gather*}
the form of Hamiltonian
\begin{equation}\label{g-ham}
\mathcal H=b_1p_{x_1}^2+\cdots +b_n p_{x_n}^2=b_1 p_{y_1}^2+\cdots+b_n p_{y_n}^2 ,
\end{equation}
and the form of induced Poisson brackets \eqref{d-br}.
\end{Proposition}

The proof is a straightforward verification of the Poisson bracket, the forms of constraints, and the form of Hamiltonian.

Below we also consider composition of $\rho_b$ (\ref{g-trans2}) and similar map $\rho_c$
\[
\rho_c\colon \  y_i= \sqrt{\frac{c_i}{\sum_{i=1}^n c_i\tilde{p}_{x_i}^2}} \tilde{p}_{x_i}\qquad\text{and}\qquad \tilde{x}_i\tilde{p}_{x_i}+y_ip_{y_i}=0 ,\qquad c_i\in \mathbb R ,
\]
which is the canonical transformation
\begin{equation}\label{bc-trans}
\sigma_{bc}\colon \ (x,p_{x})\to (\tilde{x},\tilde{p}_x)
\end{equation}
depending on $2n$ parameters $b_{i}$, $c_{i}$, $i=1,\dots,n$. This composition also preserves canonical Poisson brackets (\ref{d-br}) and the form of Hamiltonian
\[
\mathcal H=p_{x_1}^2+\cdots +p_{x_n}^2=\tilde{p}_{x_1}^2+\cdots+ \tilde{p}_{x_n}^2 .
\]
Here $\tilde{p}_{x_i}$ are momenta corresponding to coordinates $\tilde{x}_i$.

We can construct a family of equivalent integrable metrics on the sphere using these canonical transformations $\rho_b$ and $\sigma_{bc}$. For instance, applying mapping (\ref{g-trans2}) to the geodesic Hamiltonian on $T^*\mathbb S^{(n-1)}$
\begin{equation}\label{t-gen}
T=\sum_{i=1}^n a_i p_{y_i}^2+\mathcal H\sum_{i=1}^n c_iy_i^2 ,\qquad \mathcal H=\sum_{i=1}^n b_ip_{y_i}^2 ,
\end{equation}
we obtain geodesic Hamiltonian of the similar form
\begin{equation}\label{t-gen2}
\rho_b(T)=\sum_{i=1}^n b_ic_i p_{x_i}^2+\mathcal H\sum_{i=1}^n a_ib_i^{-1}x_i^2 ,\qquad \mathcal H=\sum_{i=1}^n b_ip_{x_i}^2.
\end{equation}
When $b_i=1$, we have a simple permutation of parameters $a_i\leftrightarrow c_i$ in the original Hamiltonian~(\ref{t-gen}).

This permutation of parameters is not as trivial as it seems. Let us take Hamiltonian $T$ (\ref{t-gen}) and polynomial of the second order in momenta
\begin{equation}\label{j-gen}
K=\sqrt{ w(y) } \mathcal H ,\qquad w(y)=\sum_{i\geq j} e_{ij} y_i^2y_j^2 ,
\end{equation}
where $w(y)$ is a polynomial of second order in squares $y_j^2$, which is not a full square. If we substitute $T$ (\ref{t-gen}) and $K$ (\ref{j-gen}) into
\[
\{T,K\}=0
\]
and solve the resulting system of algebraic equations for $b_i$, $c_i$, and $d_i$, we obtain a geodesic flow with two integrals of motion which are polynomials of second order in momenta.

Mapping $\rho_b$ (\ref{g-trans2}) relates second order polynomial in momenta $T$ to the second order polynomial in momenta $\rho_b(T)$ (\ref{t-gen2}) commuting with $\rho_b(K)$,
\begin{equation}\label{rk}
\{\rho_b(T),\rho_b(K)\}=0 ,\qquad \rho_b(K)=\rho\bigl(\sqrt{ w(y) } \mathcal H\bigr)=\sqrt{ \sum_{i\geq j} e_{ij}b_ib_j p^2_{x_i}p_{x_j}^2 },
\end{equation}
and with its square $\rho^2(K)$, which is a polynomial of the fourth order in momenta.

For instance, when $n=3$ and $b_i=1$, the following Hamiltonian
\begin{gather}
T=a_1p_{y_1}^2 + a_2p_{y_2}^2 + a_3p_{y_3}^2\nonumber\\
\hphantom{T=}{}-\frac{1}{2}\big((a_2+a_3)y_1^2+(a_1+a_3)y_2^2+(a_1+a_2)y_3^2 \big)\big(p_{y_1}^2+p_{y_2}^2+p_{y_3}^2\big)\label{t-trans}
\end{gather}
commutes with the polynomial of second order in momenta
\[
K=\sqrt{ w(y) } \big(p_{y_1}^2+p_{y_2}^2+p_{y_3}^2\big) ,
\]
where
\[
w(y)=\bigl((a_2 - a_3)y_1^2 + (a_3 - a_1)y_2^2-(a_1 - a_2)y_3^2\bigr)^2+4(a_3 - a_2)(a_3 - a_1)y_1^2y_2^2 .
\]
After transformation (\ref{g-trans2}), we obtain geodesic Hamiltonian~(\ref{t-gen2})
\begin{gather*}
\rho_b(T)= -\frac{1}{2}\bigl((a_2+a_3)p_{x_1}^2 + (a_1+a_3)p_{x_2}^2 + (a_1+a_2)p_{x_3}^2\bigr)\\
\hphantom{\rho_b(T)=}{}
 +\big(a_1x_1^2+a_2x_2^2+a_3x_3^2 \big)\big(p_{x_1}^2+p_{x_2}^2+p_{x_3}^2\big)
\end{gather*}
commuting with a square root $\rho_b(K)$ (\ref{rk}) and with its square $\rho^2(K)$
\begin{gather*}
 \rho^2(K)=\bigl((a_2 - a_3)p_{x_1}^2 + (a_3 - a_1)p_{x_2}^2-(a_1 - a_2)p_{x_3}^2\bigr)^2+4(a_3 - a_2)(a_3 - a_1)p_{x_1}^2p_{x_2}^2 ,
\end{gather*}
 which is the quartic polynomial in momenta.

 So, on the two-dimensional sphere, we have at least one non-trivial example of equivalent geodesic flows with quadratic and quartic polynomial invariants $T$, $K$ and $\rho_b(T)$, $\rho^2(K)$, respectively. An application of the Maupertuis principle to the construction of the corresponding nonequivalent metrics (\ref{mau-metr}) is discussed in Section~\ref{section3}.

In the next subsection, we rewrite canonical transformations $\rho_b$ (\ref{g-trans2}) and $\sigma_{bc}$ (\ref{bc-trans}) in other variables on cotangent bundle $T^*\mathbb S^2$ to the two-dimensional sphere $\mathbb S^2$ and study properties of these transformations. It allows us to construct other examples of equivalent metrics and understand how to construct similar ones on the $(n-1)$-dimensional sphere.

For brevity, below we will drop $\rho_b$ and $\sigma_{bc}$ which do not affect understanding, and simply write $H$ instead of $\rho_b(H)$ or $\sigma_{bc}(H)$.

\subsection{Euler flow on two-dimensional sphere}
The three-dimensional Euler top on the phase space $\mathfrak{so}(3)$ is defined by Hamiltonian
\begin{equation}\label{eul-ham}
\mathcal H_{\rm e}=a_1M_1^2+a_2M_2^2+a_3M_3^2
\end{equation}
commuting with any component $M_1$, $M_2$ and $M_3$ of the angular momentum vector
\[M=(M_1,M_2,M_3)\in \mathfrak{so}(3).\]
Many implicit and explicit maps preserve a form of this Hamiltonian, see~\cite{ts18e} and references within.

We consider another Hamiltonian system defined by the same Hamiltonian (\ref{eul-ham}) but on the six-dimensional phase space $T^*\mathbb S^2$, when
vector $M=x\times p_{x}$ is a cross product of two vectors $x$ and $p_x$ so that
\[
M_1=x_3p_{y_2}-x_2p_{y_3} ,\qquad M_2=x_1p_{y_3}-x_3p_{y_2} ,\qquad M_3=x_2p_1-x_1p_{y_2}.
\]
We denote the similar cross product of the vectors $y$ and $p_y$ from (\ref{g-trans2}) as $L=y\times p_y$.

By definition
\begin{alignat}{3}
& x_1^2+x_2^2+x_3^2=1 , \qquad && x_1M_1+x_2M_2+x_3M_3=0 ,&\nonumber\\
\label{caz2}
 & y_1^2+y_2^2+y_3^2=1 ,\qquad && y_1L_1+y_2L_2+y_3L_3=0 ,&
\end{alignat}
and the symplectic structure on $T^*\mathbb S^2$ is given by the bracket
\begin{alignat}{4}
& \{M_{i}, M_{j}\}=\varepsilon_{ijk}M_{k} , \qquad&&
\{M_{i}, x_{j}\} =\varepsilon_{ijk}x_{k}   \qquad&&
&\{x_{i},x_{j}\}=0 , &\nonumber\\
\label{e3}
& \{L_{i}, L_{j}\}=\varepsilon_{ijk}L_{k} ,  \qquad&&
\{L_{i}, y_{j}\} =\varepsilon_{ijk}y_{k} ,  \quad&&
&\{y_{i},y_{j}\}=0 ,&
\end{alignat}
where $\varepsilon_{ijk}$ is the skew-symmetric tensor.

Let us rewrite map (\ref{g-trans2}) in these variables on $T^*\mathbb S^2$.
\begin{Proposition}
When $b_1=b_2=b_3=1$, map $\rho_b$ \eqref{g-trans2} on $T^*\mathbb S^2$ has the following form
\begin{equation}\label{e3-rho}
 L_k=M_k ,\qquad\mbox{and}\qquad y_k^2+x_k^2+\frac{M_k^2}{M_1^2+M_2^2+M_3^2}=1 ,\qquad k=1,2,3.
 \end{equation}
This map preserves the angular momentum vector and the Poisson brackets~\eqref{e3}.
\end{Proposition}
Proof: From (\ref{g-trans2}) and (\ref{caz2}), we have
\begin{gather*}
x_1(x_3M_2 - x_2 M_3) + y_1(y_3L_2 -y_2 L_3)=0 ,\\
x_2(x_1M_3 -x_3 M_1) + y_2(y_1L_3-y_3L_1) =0 ,\\
x_3(x_2 M_1 -x_1 M_2) + y_3(y_2L_1 -y_1 L_2) =0 ,
\end{gather*}
and
\[
\frac{(y_iL_j-y_jL_i)^2}{L_1^2+L_2^2+L_3^2}=1-y_k^2-\frac{L_k^2}{L_1^2+L_2^2+L_3^2} ,\qquad i\neq j\neq k\neq i .
\]
Solving these equations for $x_i$ and $M_i$ we obtain
\begin{gather}
 x_1=\dfrac{(y_2L_3-y_3L_2)}{\sqrt{\mathcal H}} ,\qquad
M_1=\dfrac{x_2y_3(y_1L_2 -y_2 L_1)}{x_3} -\dfrac{ x_3y_2(y_3L_1 -y_1 L_3)}{x_2} ,\nonumber\\
x_2=\dfrac{(y_3L_1-y_1L_3)}{\sqrt{\mathcal H}} ,\qquad
M_2=\dfrac{ x_3y_1(y_2L_3 - y_3L_2)}{x_1}-\dfrac{x_1y_3(y_1L_2-y_2L_1)}{x_3} ,\nonumber\\
x_3=\dfrac{(y_1L_2-y_2L_1)}{\sqrt{\mathcal H}} ,\qquad
M_3=\dfrac{x_1y_2(y_3L_1 -y_1L_3)}{x_2}-\dfrac{x_2y_1(y_2L_3 -y_3L_2)}{x_1},\label{trans-m1}
\end{gather}
where
\[\mathcal H=L_1^2+L_2^2+L_3^2=M_1^2+M_2^2+M_3^2\]
is the square of the angular momentum vector. After that, we can directly verify that
\[
L_k-M_k=0 ,\qquad k=1,2,3,
\]
when constraints (\ref{caz2}) hold. Using~(\ref{trans-m1}), we can also directly check the Poisson brackets~(\ref{e3}).

\subsection{Magnetic flow on the sphere}
Let us take geodesic Hamiltonian on the sphere (\ref{g-ham})
\[
\mathcal H=a_1p_{x_1}^2+a_2p_{x_2}^2+a_3p_{x_3}^2 ,
\]
which in $(x,M)$ coordinates reads as
\begin{gather*}
\mathcal H= a_1(x_2M_3-x_3M_2)^2 +a_2(x_1M_3-x_3M_1)^2 + a_3(x_2M_1 - x_1M_2)^2  \\
\hphantom{\mathcal H}{} = a_1M_1^2 + a_2M_2^2 + a_3M_3^2 + \big(a_1x_1^2 + a_2x_2^2 + a_3x_3^2 - a_1 -a_2 - a_3\big)\big(M_1^2 + M_2^2 + M_3^2\big) .
\end{gather*}
After the shift of momenta
\[p_x\to p_x+\beta x ,\qquad \beta\in\mathbb R ,\]
we obtain the magnetic flow on the sphere defined by the Hamiltonian
\begin{equation}\label{c-ham2b}
\mathcal H=a_1(p_{x_1}+\beta x_1)^2+a_2(p_{x_2}+\beta x_2)^2+a_3(p_{x_3}+\beta x_3)^2,
\end{equation}
with linear terms in momenta \cite{bm,dor,win00}.

An integrable map preserving this flow is given by mapping (\ref{trans-m1}) after the shift of momenta.
\begin{Proposition}
Let us consider mapping on $T^*\mathbb S^2$
\[\rho_\beta\colon \  (y,L)\to (x,M)\]
 defined as
\begin{gather*}
x_1=\dfrac{(y_2L_3-y_3L_2+\beta y_1)}{\sqrt{\mathcal H}} ,\qquad\!\!
x_2=\dfrac{(y_3L_1-y_1L_3+\beta y_3)}{\sqrt{\mathcal H}} ,\qquad\!\!
x_3=\dfrac{(y_1L_2-y_2L_1+\beta y_3)}{\sqrt{\mathcal H}} ,
\end{gather*}
and
\begin{gather}
M_1=\dfrac{x_2y_3(y_1L_2 -y_2 L_1+\beta y_3)}{x_3} -\dfrac{ x_3y_2(y_3L_1 -y_1 L_3+\beta y_2)}{x_2} ,\nonumber\\
M_2=\dfrac{ x_3y_1(y_2L_3 - y_3L_2+\beta y_1)}{x_1}-\dfrac{x_1y_3(y_1L_2-y_2L_1+\beta y_3)}{x_3} ,\nonumber\\
M_3=\dfrac{x_1y_2(y_3L_1 -y_1L_3+\beta y_2)}{x_2}-\dfrac{x_2y_1(y_2L_3 -y_3L_2+\beta y_1)}{x_1} .\label{trans-2b}
\end{gather}
This mapping preserves the form of Hamiltonian $\mathcal H$ \eqref{c-ham2b}, Poisson bracket \eqref{e3} and values of its Casimir functions \eqref{caz2}.
\end{Proposition}
In contrast with the previous transformation~(\ref{e3-rho}), this transformation changes the angular momentum vector so that $L_i\neq M_i$.

Thus, we rewrite map $\rho_b$ (\ref{g-trans2}) using the entries of the angular momentum vector (\ref{trans-m1}) and construct its trivial generalisation $\rho_\beta$ (\ref{trans-2b}). Below we apply these maps to construct equivalent geodesic and non-equivalent potential flows on the two-dimensional sphere.

\section{Main example of equivalent metrics on the sphere}\label{section3}

Elliptic coordinate system $u_1,\dots, u_{n-1}$ on the sphere
 with parameters $a_1<\cdots<a_n$ is defined through equation
\begin{equation}\label{ell-sph}
\sum_{i=1}^{n}\frac{x_i^2}{\lambda-a_i}=\frac{\prod_{k=1}^{n-1}(\lambda-u_k)}{\prod_{i=1}^n(\lambda-a_i)} ,
\end{equation}
that implies $\sum x_i^2=1$. Elliptic coordinates are orthogonal and locally defined, they take values in the intervals
\[a_1<u_1<a_2<u_2<\cdots< u_{n-1}<a_n .\]
The Poisson bracket between elliptic coordinates $u_{k}$ and their conjugated momenta $p_{u_{k}}$ is the canonical Poisson bracket
\[
\{u_i,u_j\}= 0 ,\qquad \{p_{u_i},p_{u_j}\}=0 ,\qquad \{u_i,p_{u_j}\}=\delta_{ij}  ,\qquad i,j=1,\dots,n-1.
\]
When $n=3$ six variables $x_i$ and $M_i$ are expressed via four variables $u_{1,2}$ and $p_{u_{1,2}}$ in the following way
\begin{gather}
x_i=\sqrt{\dfrac{(u_1-a_i)(u_2-a_i)}{(a_j-a_i)(a_k-a_i)}}
,\qquad i\neq j\neq k\neq i ,\nonumber\\
\label{xm-ell}
M_i=\dfrac{2\varepsilon_{ijk}x_jx_k(a_j-a_k)}{u_1-u_2}
\bigl((a_i-u_1)p_{u_1}-(a_i-u_2)p_{u_2}\bigr) .
\end{gather}
Similar second pair of elliptic coordinates $v_{1,2}$ on the sphere together with the conjugated momenta $p_{v_{1,2}}$
\[
\{v_1,v_2\}=\{v_1,p_{v_2}\}=\{v_2,p_{v_1}\}=\{p_{v_1},p_{v_2}\}=0 ,\qquad \{v_1,p_{v_1}\}=\{v_2,p_{v_2}\}=1 ,
\]
determine the second set of variables on $T^*\mathbb S^2$
\begin{gather*}
y_i=\sqrt{\dfrac{(v_1-a_i)(v_2-a_i)}{(a_j-a_i)(a_k-a_i)}},\qquad i\neq j\neq k\neq i ,\\
%\label{ym-ell}\\
L_i=\dfrac{2\varepsilon_{ijk}y_jy_k(a_j-a_k)}{v_1-v_2}
\bigl((a_i-v_1)p_{v_1}-(a_i-v_2)p_{v_2}\bigr) .
\end{gather*}
In elliptic coordinates, the square of the angular momentum is equal to
\begin{equation}\label{eul-ham1}
\mathcal H=\frac{4(a_1-u_1)(a_2-u_1)(a_3 - u_1)p_{u_1}^2}{u_1-u_2}+\frac{ 4(a_1 - u_2)(a_2 - u_2)(a_3 - u_3)p_{u_1}^2}{u_2 - u_1}
\end{equation}
and mapping (\ref{trans-m1})
 \[\rho_{b}\colon \ (u,p_u)\to(v,p_v)\]
 can be rewritten using a pair of the so-called Abel polynomials on auxiliary variable $z$ defining intersection divisor on the hyperelliptic curve~$C$ \cite{ts17v,ts17c}:
\begin{gather}
\mathcal P(z)=\frac{(z - v_2)\varphi(v_1)p_{v_1}}{v_1 -v_2} + \frac{(z-v_1)\varphi(v_2)p_{v_2}}{v_2-v_1}\nonumber\\
\hphantom{\mathcal P(z)}{} =-
\frac{(z - u_2)\varphi(u_1)p_{u_1}}{u_1 - u_2} - \frac{(z-u_1)\varphi(u_2)p_{u_2}}{u_2-u_1}\label{pz}
\end{gather}
and
\begin{equation} \label{psi}
\psi(z)=\mathcal H(z-v_1)(z-v_2)(z-u_1)(z-u_2)=\varphi(z)(z\mathcal H+\mathcal H_{\rm e})-\mathcal P(z)^2 ,
\end{equation}
where $ \varphi(z)=-(a_1-z)(a_2-z)(a_3-z)$, $\mathcal H$ is the square of the angular momentum vector (\ref{eul-ham1}) and $\mathcal H_{\rm e}$ is the Hamiltonian of the Euler top (\ref{eul-ham})
\begin{gather*}
\mathcal H_{\rm e}= a_1M_1^2+a_2M_2^2+a_3M_3^2=\dfrac{4u_2\varphi(u_1)p_{u_1}^2}{u_1 - u_2} +\dfrac{4u_1\varphi(u_2)p_{u_2}^2}{u_2-u_1}\\
\hphantom{\mathcal H_e}{}
= a_1L_1^2+a_2L_2^2+a_3L_3^2=\dfrac{4v_2\varphi(v_1)p_{v_1}^2}{v_1 - v_2} +\dfrac{4v_1\varphi(v_2)p_{v_2}^2}{v_2-v_1} .
\end{gather*}
These equations (\ref{pz}) and (\ref{psi}) should be interpreted as an identity for $z$ and each set of elliptic variables $u$, $p_u$ and $v$, $p_v$.

Transformation $\rho_b$ (\ref{e3-rho}) is a partial case of the integrable maps associated with the nonholonomic Chaplygin and Veselova systems on the sphere \cite{ts17v,ts17c}.

\subsection{Integration of the original integrable flow}
Let us come back to the geodesic Hamiltonian (\ref{t-trans}), which becomes additive separable Hamiltonian in elliptic variables
\begin{equation}\label{ham-sep-m}
H=\mathcal{S}_1+\mathcal{S}_2 ,\qquad\mathcal{S}_k=-(a_1-v_k)(a_2-v_k)(a_3-v_k)p_{v_k}^2 ,
\end{equation}
commuting with linear integrals of motion
\[
I_1=\sqrt{(a_1-v_1)(a_2-v_1)(a_3-v_1)} p_{v_1}\qquad\text{and}\qquad I_2=\sqrt{(a_1-v_2)(a_2-v_2)(a_3-v_2)} p_{v_2} ,
\]
quadratic integral of motion $J=\mathcal{S}_1-\mathcal{S}_2$ (\ref{j-gen}) and any other functions $f(\mathcal{S}_1,\mathcal{S}_2)$ on~$\mathcal{S}_{1,2}$.

The corresponding diagonal metric
\begin{equation}\label{varphi}
\mathrm g(v_1,v_2)=\left(
 \begin{matrix}
 \varphi( v_1) & 0 \\
 0 & \varphi(v_2)
 \end{matrix}
 \right) ,\qquad \varphi(z)=-(a_1-z)(a_2-z)(a_3-z)
\end{equation}
has a non-trivial isometry group. Integrals of motion $f(\mathcal{S}_1,\mathcal{S}_2)$ are in involution with respect to the Poisson brackets associated with the canonical Poisson bivector $P$,
\begin{equation}\label{p1}
P=\dfrac{\partial}{\partial v_1}\wedge\frac{\partial}{\partial p_{v_1}}+\dfrac{\partial}{\partial v_2}\wedge\frac{\partial}{\partial p_{v_2}}
\end{equation}
 and second compatible Poisson bivector
 \begin{equation}\label{p2}
P'=\mathcal{S}_1 \dfrac{\partial}{\partial v_1}\wedge\frac{\partial}{\partial p_{v_1}}+\mathcal{S}_2 \dfrac{\partial}{\partial v_2}\wedge\frac{\partial}{\partial p_{v_2}} .
\end{equation}
This pair of compatible Poisson bivectors determines the bi-Hamiltonian vector field
\[
X=P'{\rm d}H=P{\rm d}H' ,\qquad\text{where}\quad H'=\frac{\mathcal{S}_1^2+\mathcal{S}_2^2}{2} .
\]
The Hamilton--Jacobi equation $H=E$ (\ref{ham-sep-m}) admits additive separation
\[
H=E_1+E_2 ,\qquad \mathcal{S}_1(v_1,p_{v_1})=E_1, \qquad \mathcal{S}_2(v_2,p_{v_2})=E_2 .
\]
Because
\begin{equation}\label{meq-sep}
\dfrac{d{v}_k}{dt}=\{H,v_k\}=2(a_1 - v_k)(a_2 - v_k)(a_3 - v_k)p_{v_k} ,\qquad k=1,2,
\end{equation}
and
\begin{equation}\label{p-ell}
p_{v_k}^2=-\frac{E_k}{(a_1 - v_k)(a_2 - v_k)(a_3 - v_k)} ,
\end{equation}
we have the following separate equations
\begin{equation}\label{sep-flow}
\left(\dfrac{{\rm d}{v}_k}{{\rm d}t}\right)^2+4(a_1 - v_k)(a_2 - v_k)(a_3 - v_k)E_k=0 ,\qquad k=1,2.
\end{equation}
Standard substitution
\begin{equation}\label{v12}
v_{k}=\dfrac{a_1+a_2+a_3}{3} +\dfrac{w}{E_k} ,\qquad \dfrac{{\rm d}v_1}{{\rm d}t}=\dfrac{1}{E_k}\dfrac{{\rm d}w}{{\rm d}t} ,
\end{equation}
reduces equations (\ref{meq-sep}) to equations for the elliptic Weierstrass function
\[
\left(\frac{{\rm d}w}{{\rm d}t}\right)^2=4w^3-g_2w+g_3 ,
\]
where
\begin{gather*}
g_2=\dfrac{4(a_1^2 - a_1a_2 - a_1a_3 + a_2^2 - a_2a_3 + a_3^2)E_k^2}{3} ,\\
g_3=\dfrac{4(2a_1 - a_2 - a_3)(a_1 -2 a_2+a_3)(a_1 +a_2 -2 a_3)E_k^3}{27} .
\end{gather*}
Thus, we can express variables $v_{1}$, $v_{2}$ (\ref{v12}) and $p_{v_1}$, $p_{v_2}$ (\ref{p-ell}) via two elliptic $\wp$-functions on time.

Below we apply transformation $\rho_b$ (\ref{e3-rho}) to this simple geodesic flow~(\ref{sep-flow}).

\subsection{Some properties of the equivalent metrics}
Let us introduce the diagonal matrix
\[
\mathbf A=\left(
 \begin{matrix}
 {a}_1 & 0 & 0 \\
 0 & {a}_2 & 0 \\
 0 & 0 & {a}_3
 \end{matrix}
 \right) ,
\]
and its spectral characteristics
\begin{equation}\label{bcd}
b=a_1+a_2+a_3 ,\qquad c=a_1a_2+a_1a_3+a_2a_3 ,\qquad d=a_1a_2a_3.
\end{equation}
It allows us to rewrite Hamiltonian $H$ (\ref{ham-sep-m}) using the angular momentum vector
\[
H=\dfrac{1}{2}(L,\mathbf A L) +\dfrac{(x,\mathbf A x) - b}{4} (L,L) .
\]
After transformation $\rho_b$ (\ref{e3-rho}) this Hamiltonian has the following form
\begin{equation}\label{Hams}
H=\frac{1}{4}(M,\mathbf AM)-\frac{1}{4}(x,\mathbf Ax)(M,M) ,
\end{equation}
so in elliptic coordinates (\ref{xm-ell}) we have
\begin{equation}\label{H}
H= \mathrm g_{11} p_{u_1}^2+\mathrm g_{22}p_{u_2}^2=\frac{(u_1+2u_2-b)\varphi(u_1)p_{u_1}^2}{u_1-u_2}+\frac{( 2u_1 + u_2-b)\varphi(u_2)p_{u_2}^2}{u_2 - u_1} ,
\end{equation}
which allows us to calculate the corresponding diagonal metric on $\mathbb S^2$
\begin{equation}\label{metr}
\mathrm g(u_1,u_2)=\left( \begin{matrix}
 \dfrac{(u_1+2u_2-b)\varphi(u_1)}{u_1-u_2} &0 \\
 0 &\dfrac{( 2u_1 + u_2-b)\varphi(u_2)}{u_2 - u_1}
 \end{matrix} \right) .
\end{equation}
For metric space $\big(\mathbb S^2,\mathrm g\big) $ we can define a vector space of symmetric $ (m, 0)$ Killing tensors $\mathrm K$, which are solutions of the Killing equations
\begin{equation}\label{kill-eq}
[\![\mathrm g, \mathrm K]\!]=0 ,
\end{equation}
where $[\![\cdot,\cdot]\!]$ is a Schouten bracket.

When $m = 1$, solutions $\mathrm K$ are said to be infinitesimal isometries that form an isometry group. According to~\cite{grom99}: ``everybody knows that isometry group ${\rm Isom}(M, g) = {\rm Id}$ for generic Riemannian or pseudo-Riemannian metrics $\mathrm g $ for $\dim M\geq 2$''. Our metric is no exception.
\begin{Proposition}
Metric $\mathrm g$ \eqref{metr} on a two-dimensional sphere~$\mathbb S^2$ has the trivial isometry group and trivial vector spaces of Killing tensors of valency two and three when $m=2,3$.
\end{Proposition}
\par\noindent
The proof is a straightforward solution of the Killing equation (\ref{kill-eq}) at $m=1,2,3$.

Transformation $\rho_b$ (\ref{e3-rho}) maps integral of motion
\[K=\mathcal{S}_1 \mathcal{S}_2\]
to the following polynomial of fourth order in momenta
\[K =K_1^2\cdot K_2=\dfrac{1}{16}(M,\mathbf Ax)^2(M,M) .\]
In elliptic coordinates (\ref{xm-ell}) these factors are equal to
\begin{equation}\label{K}
 K_1=\frac{p_{u_1}-p_{u_2}}{u_1-u_2} , \qquad K_2=\frac{\varphi(u_1)\varphi(u_2)\Bigl(\varphi(u_1)p_{u_1}^2-\varphi(u_2)p_{u_2}^2\Bigr)}{u_1-u_2} ,
\end{equation}
which allows us to determine the Killing tensor $\mathrm K$ of valency $m=4$ on the sphere $\mathbb S^2$, which satisfies the Killing equation (\ref{kill-eq}).

Transformation $\rho_b$ (\ref{e3-rho}) preserves the form of the canonical Poisson bivector $P$ (\ref{p1})
\begin{equation}\label{pp1}
P=\left(
 \begin{array}{cccc}
 0 & 0 & 1 & 0 \\
 0 & 0 & 0 & 1 \\
 -1 & 0 & 0 & 0 \\
 0 & -1 & 0 & 0 \\
 \end{array}
 \right)
\end{equation}
and changes the form of the second Poisson bivector $P'$ (\ref{p2})
\begin{equation}\label{pp2}
P'=\left(
 \begin{matrix}
 0 & A_1 & A_2 & A_3 \\
 -A_1 & 0 & A_4 & A_5 \\
 -A_2 & -A_4 & 0 & A_6 \\
 -A_3 & -A_5 & -A_6 & 0
 \end{matrix}
 \right),
\end{equation}
where
\begin{gather*}
A_1=\frac{2\varphi_1\varphi_2(p_{u_1} - p_{u_2})}{(u_1 - u_2)^2} ,\qquad  A_2=-\mathrm g_{11}p_{u_1}^2 -\frac{\varphi_1\varphi_2(p_{u_1} - p_{u_2})^2}{(u_1 - u_2)^3} ,  \\
A_5=-\mathrm g_{22}p_{u_2}^2+\frac{\varphi_1\varphi_2(p_{u_1} - p_{u_2})^2}{(u_1 - u_2)^3} ,
\qquad
A_3=\frac{\varphi_1^2(p_{u_1} - p_{u_2})^2}{(u_1 - u_2)^3}-\mathrm g_{11}p_{u_2}(2p_{u_1} - p_{u_2}) ,\\
 A_4=-\frac{\varphi_2^2(p_{u_1} - p_{u_2})^2}{(u_1 - u_2)^3}+\mathrm g_{22}p_{u_1}(p_{u_1} - 2p_{u_2}),
 \\
A_6=-\frac{(p_{u_1} - p_{u_2})p_{u_1}p_{u_2}}{(u_1 - u_2)^2}\left( \frac{\partial}{\partial u_1}\frac{\varphi_1\varphi_2}{(u_1-u_2)^3}-\frac{\partial}{\partial u_2} \left(\mathrm g_{22}-\frac{\varphi_2^2}{(u_1-u_2)^3}\right) \right) ,
\end{gather*}
and $\varphi_k=\varphi(u_k)$ for brevity. The structure of this Poisson bivector $P'$ is completely different from the structure of the so-called natural Poisson bivectors on the sphere \cite{ts11,vt09}.

Both bivectors $P$ (\ref{p1}) and $P'$ (\ref{pp2}) are invertible, which allows us to introduce a hereditary or recursion operator defined as
\[N=P' P^{-1} .\]
The spectral curve of $N$ has the form
\[
\det(N-\lambda {\rm Id})=\big(\lambda^2+H\lambda-K\big)^2 ,
\]
where $H$ and $K$ are given by (\ref{H})--(\ref{K}). Thus, the following equation holds
\begin{equation*}%\label{j-ham}
P'{\rm d}H=P{\rm d}J ,\qquad J=\frac{H^2}{2}-K=\frac{\mathcal{S}_1^2+\mathcal{S}_2^2}{2},
\end{equation*}
 where $P$ and $P'$ are given by (\ref{pp1}) and (\ref{pp2}) and
\[
{\rm d}H=\left(
 \begin{matrix}
 \dfrac{\partial H}{\partial u_1} \vspace{1mm}\\
 \dfrac{\partial H}{\partial u_2} \vspace{1mm}\\
 \dfrac{\partial H}{\partial p_{u_1}} \vspace{1mm}\\
 \dfrac{\partial H}{\partial p_{u_2}}
 \end{matrix}
 \right)\qquad\mbox{and}\qquad
{\rm d}H'=\left(
 \begin{matrix}
 \dfrac{\partial H'}{\partial u_1} \vspace{1mm}\\
 \dfrac{\partial H'}{\partial u_2} \vspace{1mm}\\
 \dfrac{\partial H'}{\partial p_{u_1}} \vspace{1mm}\\
 \dfrac{\partial H'}{\partial p_{u_2}}
 \end{matrix}
 \right) .
\]
So, we have a geodesic flow on the bi-Hamiltonian manifold $\big(T^*\mathbb S^2, P, P'\big)$ defined by the coefficients of the Casimir functions of the Poisson pencil $P_\lambda=P+\lambda P'$.
\begin{Proposition}
On the cotangent bundle $T^*\mathbb S^2$ Hamiltonian $H$ \eqref{H}
\begin{gather*}
H=\frac{1}{4}(M,\mathbf AM)-\frac{1}{4}(x,\mathbf Ax)(M,M)
=\frac{(u_1+2u_2-b)\varphi(u_1)p_{u_1}^2}{u_1-u_2}+\frac{( 2u_1 + u_2-b)\varphi(u_2)p_{u_2}^2}{u_2 - u_1}
\end{gather*}
yields bi-Hamiltonian vector field
\[
X=P{\rm d}H=\bigl(N^{-1}P\bigr) {\rm d}J , \qquad N=P' P^{-1},
\]
where $P$ and $P'$ are given by~\eqref{pp1} and~\eqref{pp2}.
\end{Proposition}

The proof is a straightforward calculation.

\subsection{Potential motion}
Let us discuss potentials $V(x)$ which can be added to the geodesic Hamiltonian~$H$~(\ref{Hams}).
For instance, starting with the following separable Hamiltonian
\[
 H_1=\mathcal{S}_1+\mathcal{S}_2+\beta \frac{v_1\mathcal{S}_1-v_2\mathcal{S}_2}{\mathcal{S}_1-\mathcal{S}_2},
\]
so that
\[
H_1=E\ \Rightarrow\ \mathcal{S}_1^2+\beta v_1\mathcal{S}_1-E\mathcal{S}_1=\mathcal{S}_2^2+\beta v_2\mathcal{S}_2-E\mathcal{S}_2 ,
\]
we obtain Hamiltonian
\begin{gather*}
H_1=(M,\mathbf AM)-(x,\mathbf Ax)(M,M)+V(x) ,\qquad\!
V(x)=\beta(\mathbf A x,x)=\beta \big(a_1x_1^2+a_2x_2^2+a_3x_3^2\big) ,
\end{gather*}
commuting with the second integral of motion
\[
H_2= (M,\mathbf Ax)^2\bigl((M,M)-\beta\bigr).
\]
Other potentials may be obtained using the following substitution
\begin{gather}
H_1=(M,\mathbf AM)-(x,\mathbf Ax)(M,M)+V(x) ,\nonumber\\
H_2=\bigl( (M,\mathbf Ax)^2+(M,\mathbf Ax) U_1(x)+U_2(x)\bigr)\bigl((M,M)+U_3(x)\bigr) .\label{ham-pot1}
\end{gather}
The equation $\{H_1, H_2\}=0$ has several solutions from which we single out the following polynomial ``cubic'' potential
\[
V(x)=\alpha\left((x,\mathbf Ax) - \dfrac{b}{3}\right)^3 ,\qquad \alpha\in\mathbb R ,
\]
where $b=a_1+a_2+a_3$, and
\begin{gather*}
U_1(x)=0 ,\qquad U_2(x)=0 ,\qquad U_3(x)=-\alpha W(x) ,
\\
W(x)=(a_1 - a_3)(a_1 - a_2)x_1^4+(a_2-a_1)(a_2-a_3)x_2^4+(a_3-a_1)(a_3-a_2)x_3^4\\
\hphantom{W(x)=}{}
-\big(a_1^2 + 2a_2a_3\big)x_1^2-\big(a_2^2+2a_1a_3\big)x_2^2-\big(a_3^2+2a_1a_2\big)x_3^2+\dfrac{b^2}{3} .
\end{gather*}
In this case polynomial $H_1$ (\ref{ham-pot1}) becomes a rational function in $v$-variables
\[
H_1=(\mathcal{S}_1+\mathcal{S}_2)+\alpha\left(\frac{(3v_1-b)\mathcal{S}_1-(3v_2-b)\mathcal{S}_2}{\mathcal{S}_1-\mathcal{S}_2}\right)^3
\]
and we know nothing about the separation of variables and the bi-Hamiltonian structure of the corresponding vector field when $\alpha\neq 0 $.

Following \cite{bol95, ts99,ts01}, we can use the Maupertuis principle to construct a new metric on the sphere associated with the Hamiltonian
\[
\widetilde{H}=\frac{\mathrm g_{11}p^2_{u_1}+\mathrm g_{22}p^2_{u_2}}{E-V(x)} .
\]
The corresponding additional integral of motion is the polynomial of the fourth order in momenta.

\section{One family of equivalent metrics}

As an example, we take another separable Hamiltonian
\[
H=\hat{\lambda}_1+\hat{\lambda}_2 ,\qquad \hat{\lambda}_k=v_k\mathcal{S}_k=-v_k(a_1-v_k)(a_2-v_k)(a_3-v_k)p_{v_k}^2 ,
\]
which in $(y,L)$ variables reads as
\[H=a_1(y_2L_3-y_3L_2)^2 + a_2(y_1L_3-y_3L_1)^2 + a_3(y_2L_1 -y_1L_2)^2 ,\]
up to the factor $1/4$. In Cartesian coordinates, it has the form
\[
\hat{H}=a_1p_{y_1}^2+a_2p_{y_2}^2+a_3p_{y_3}^2 .
\]
Below we apply transformations $\rho_b$~(\ref{g-trans2}) with different values of $b_i$ to this Hamiltonian and present equivalent metrics on the sphere related to canonical transformation~$\sigma$~(\ref{bc-trans}).

\textbf{Case 1.} Using transformation $\rho_b$ (\ref{g-trans2}) with
\[b_1=b_2=b_3=1,\]
we obtain the following integrals of motion
\begin{gather}
H^{(1)}= \left(\dfrac{2\varphi(u_1)}{(u_1-u_2)^2} - u_1 +\dfrac{(b^2 - bu_2 - 2c}{u_1-u_2}\right)\varphi_1p_{u_1}^2-\dfrac{4\varphi_1\varphi_2 p_{u_1}p_{u_2}}{(u_1-u_2 )^2}
\nonumber \\
\hphantom{H^{(1)}=}{}  + \left(\dfrac{2\varphi(u_2)}{(u_1 -u_2)^2} - u_2 +\dfrac{b^2 - bu_1 - 2c}{u_2-u_1}\right)\varphi_2p_{u_2}^2,\label{case1} \\
K^{(1)}= \varphi(u_1)\varphi(u_1) K_1^2 K_2 ,\qquad\text{where}\quad K_1=\dfrac{p_1-p_{y_2}}{u_1-u_2} ,\nonumber
\end{gather}
where
\begin{gather*}
K_2= \left(\dfrac{\varphi(u_2)}{(u_1-u_2)^2} - u_2 - \frac{u_2(b - 2u_1)}{u_1-u_2}\right)\varphi(u_1)p_{u_1^2}
 -\dfrac{ 2\varphi(u_1)\varphi(u_2)p_{u_1}p_{u_2}}{(u_1-u_2)^2} \\
 \hphantom{K_2=}{}
 +  \left(\dfrac{\varphi(u_1)}{(u_1-u_2)^2} - u_1 -\frac{u_1(b - 2u_2)}{u_2-u_1}\right)\varphi(u_2)p_{u_2}^2 ,
\end{gather*}
and $\varphi(z)$ is given by (\ref{varphi}).

\textbf{Case 2.} Using transformation $\rho_b$ (\ref{g-trans2}) with
\[b_1=a_1,\qquad b_2=a_2,\qquad b_3=a_3,\]
 we obtain the following integrals of motion
 \begin{gather}
 H^{(2)}=
 \dfrac{\bigl(d - cu_1 + (u_1 - 2u_2 + b)u_1u_2\bigr)\varphi(u_1)p_{u_1}^2}{(u_1-u_2)^2}
 + \dfrac{ 2\varphi(u_1)\varphi(u_2)p_{u_1}p_{u_2}}{(u_1-u_2)^2}\nonumber\\
\hphantom{H^{(2)}=}{}
 + \dfrac{\bigl(d - cu_2 + (u_2 - 2u_1 + b)u_1u_2\bigr)\varphi(u_2)p_{u_2}^2}{(u_1-u_2)^2} ,\nonumber\\
 K^{(2)}=K_1^2K_2=\left( \dfrac{u_1p_{u_1}-u_2p_{u_2}}{u_1-u_2}\right)^2\cdot \dfrac{\varphi(u_1)\varphi(u_2)\bigl(\varphi(u_1)p_{u_1}^2 - \varphi(u_2)p_{u_2}^2\bigr)}{u_1u_2(u_1 - u_2)} . \label{case2}
\end{gather}

\textbf{Case 3.} Using transformation $\rho_b$ (\ref{g-trans2}) with
\[b_1=a_1^2,\qquad b_2=a_2^2,\qquad b_3=a_3^2,\]
 we obtain the following integrals of motion
\begin{gather}
H^{(3)}= \dfrac{u_1\bigl( 2u_1+u_2-\tilde{b}u_1 u_2\bigr)\varphi(u_1)p_1^2}{u_1-u_2}+
\dfrac{u_2\bigl( u_1+2u_2-\tilde{b}u_1u_2 \bigr)\varphi(u_2)p_{y_2}^2}{u_2-u_1},\nonumber\\
K^{(3)}=K_1^2K_2=\left(\dfrac{u_1^2p_{1}-u_2^2p_{2}}{u_1-u_2}\right)^2\cdot
\dfrac{\varphi(u_1)\varphi(u_2)\bigl(u_1\varphi(u_1)p_{1}^2-u_2\varphi(u_2)p_{2}^2\bigr)}{u_1-u_2} , \label{case3}
\end{gather}
where $\tilde{b}=a_1^{-1}+a_2^{-1}+a_3^{-1}$.

\begin{Proposition}
Geodesic Hamiltonians $H^{(1)}$ \eqref{case1}, $H^{(2)}$ \eqref{case2} and $H^{(3)}$ \eqref{case3} are related to each other by canonical transformations~$\sigma$ \eqref{bc-trans} with the suitable set of parameters.
\end{Proposition}
Thus, we have three equivalent metrics on the two-dimensional sphere.

\subsection{Potential motion}
We can try to destroy this equivalence of the geodesic flows by adding potentials to the geodesic Hamiltonian, for instance,
changing Hamiltonians in the following way
\[H_1=H+V\qquad\mbox{and}\qquad H_2=K_1^2(K_2+U) .\]
Let us present some potentials explicitly:
\begin{gather*}
V^{(1)}=\dfrac{1}{b-u_1-u_2} \left( \alpha\left(b-\dfrac{c}{b-u_1-u_2} -\dfrac{d}{(b-u_1-u_2)^2}\right)+\beta\right) ,\\
U^{(1)}=\dfrac{\varphi(u_1)\varphi(u_2)}{b-u_1-u_2} \left(\alpha \dfrac{u_2\varphi(u_1)-u_1\varphi(u_2)}{(b - u_1 - u_2)^2(u_1-u_2)}+\beta\right) ,
\end{gather*}
and
\begin{gather*}
V^{(2)}=u_1u_2\big(\alpha\bigl(u_1^2u_2^2-cu_1u_2+bd\big)+\beta\big) ,\\
U^{(2)}=\varphi(u_1)\varphi(u_2)\big(\alpha\big(u_1^2u_2^2 - cu_1u_2 + d(u_1 + u_2)\big)+\beta\big) .
\end{gather*}
In the third case, we have
\begin{gather*}
V^{(3)}= \alpha \dfrac{(u_1 + u_2)(u_1 +u_2 +\tilde{b} u_1u_2)^2}{u_1^3u_2^3}+\beta \dfrac{ u_1 + u_2}{u_1u_2} ,
\\
U^{(3)}=  \dfrac{\varphi(u_1)\varphi(u_2)}{4u_1u_2(u_1-u_2)^2}
\left(\alpha
\dfrac{d(u_1^2 + u_1u_2 + u_2^2) -c u_1u_2(u_1 + u_2)}{u_1^2u_2^2}
+\beta
\right) .
\end{gather*}
Using relations
\[
u_1+u_2=a_1x_1^2+a_2x_2^2+a_3x_3^2-b ,\qquad u_1u_2=a_2a_3x_1^2+a_1a_3x_2^2+a_1a_2x_3^2,\]
we can rewrite these potentials in terms of Cartesian coordinates.

As above, the Maupertuis principle allows the construction of a new metric on the sphere associated with the Hamiltonian
\[
\widetilde{H}=\frac{\mathrm g_{11}p^2_{u_1}+\mathrm g_{22}p^2_{u_2}}{E-V},
\]
where $\mathrm g$ and $V$ are metrics and potentials associated with $H^{(1)}$ (\ref{case1}), $H^{(2)}$ (\ref{case2}) and $H^{(3)}$ (\ref{case3}).

\subsection{Hamiltonians with linear in momenta terms}
Let us come back to the separable Hamiltonian (\ref{ham-sep-m})
\[
H=\mathcal{S}_1+\mathcal{S}_2=\varphi(v_1)p_{v_1}^2+\varphi(v_2)p_{v_2}^2 .
\]
 Using transformation $\rho_b$ (\ref{g-trans2}) with
\[b_1=a_1,\qquad b_2=a_2,\qquad b_3=a_3,\]
we obtain
\begin{gather*}
H= \left(2a_1 - \dfrac{a_2 + a_3}{a_2a_3} w\right)(x_2M_3-x_3M_2)^2
+\left(2a_2 -\dfrac{a_1 + a_3}{a_1a_3} w\right)(x_1M_3-x_3M_1)^2\\
\hphantom{H=}{} + \left(2a_3 -\dfrac{a_1 + a_2}{a_1a_2} w\right)(x_2M_1 -x_1M_2)^2 ,
\end{gather*}
where
\[
w=u_1u_2=a_2a_3x_1^2+a_1a_3x_2^2+a_1a_2x_3^2 .
\]
In elliptic coordinates, this Hamiltonian has the following form
\begin{equation}\label{b-h0}
H={\mathrm g}_{11}p_{u_1}^2+2{\mathrm g}_{12}p_{u_1}p_{u_2}+{\mathrm g}_{22}p_{u_2}^2 ,
\end{equation}
where the metric is
\[
{\mathrm g}=\left(
 \begin{matrix}
 \dfrac{d(2u_1 - u_2) - u_1^2u_2^2 + bu_1u_2^2 - cu_1u_2}{(u_1 - u_2)^2} & \dfrac{u_1u_2\varphi(u_1)\varphi(u_2)}{(u_1 - u_2)^2}\vspace{2mm} \\
 \dfrac{u_1u_2\varphi(u_1)\varphi(u_2)}{(u_1 - u_2)^2} &\dfrac{d(u_1 -2 u_2) - u_1^2u_2^2 + bu_1^2u_2 - cu_1u_2}{(u_1 - u_2)^2}
 \end{matrix}
 \right) ,
\]
and $b$, $c$ and $d$ are combinations of $a_1$, $a_2$ and $a_3$~(\ref{bcd}). The corresponding quartic invariant is a product of two polynomials in momenta
\begin{equation}\label{b-k0}
K=\left(\frac{u_1p_{u_1} -u_2p_{u_2} }{u_1 - u_2}\right)^2\frac{\varphi(u_1)\varphi(u_2)\big(u_1\varphi(u_1)p_{u_1}^2-u_2\varphi(u_2)p_{u_2}\big)}{u_1-u_2} .
\end{equation}
These integrals of motion $H$ and $K$ coincide with integrals $H$ (\ref{H}) and $K$ (\ref{K}) after canonical transformation $\sigma_{bc}$ (\ref{bc-trans}) depending on parameters $b_i=1$ and $c_k=a_k$.

The main difference is that canonical transformation $p_k\to p_k+\beta_k$ acts trivially when $b_i=1$
\[
\mathcal H=\sum (p_k+\beta x_k)^2=\sum p_k^2+2\beta\sum x_kp_k+\beta^2\sum x_k^2=\sum p_k^2+\beta^2
\]
according to constraints (\ref{f-con}). When $b_i=a_i$, this transformation adds nontrivial term to the Hamiltonian $\mathcal H$ (\ref{g-ham}), which is linear polynomial in momenta
\[
\mathcal H=\sum a_k (p_k+\beta x_k)^2=\sum a_k p_k^2+2\beta\sum a_kx_kp_k+\beta^2\sum a_kx_k^2 .
\]
As a result, applying a transformation (\ref{trans-2b}) to the Hamiltonian
\[
\tilde{H}=\tilde{\mathcal S}_1+\tilde{\mathcal{S}}_2 ,\qquad \tilde{\mathcal S}_k=\mathcal{S}_k+\beta^2 v_k=\varphi(v_k)p_{v_k}^2+\beta^2 v_k ,
\]
we obtain Hamiltonian on the $T^*\mathbb S^2$
\[
\tilde{H}=H-\frac{2\beta u_1u_2\big(u_2\varphi(u_1)p_{u_1} -u_1\varphi(u_2)p_{u_2}\big)}{u_1 - u_2}
-\beta^2\left(\frac{u_2^2\varphi(u_1)-u_1^2\varphi(u_2)}{u_1 - u_2} +d(u_1 + u_2)\right)
\]
involving linear terms in velocity. Here $H$ is given by (\ref{b-h0}) and the corresponding second integral of motion is equal to
\[
\tilde{K}=K+\beta^4 B_4+\beta_3B_3+\beta_2B_2+\beta B_1,
\]
where $K$ is given by (\ref{b-k0}) and
\begin{gather*}
B_4=-bu_1^2u_2^2 -d\big(u_1^2 + 2u_1u_2 + u_2^2\big) + cu_1u_2(u_1 + u_2) ,
\\
B_3=-\dfrac{2u_1u_2\big(u_2^2\varphi(u_1)p_{u_1} -u_1^2\varphi(u_1)p_{u_2}\big)}{u_1 - u_2} ,
\\
B_1=-\dfrac{2u_1u_2\varphi(u_1)\varphi(u_2)p_{u_1}p_{u_2}(u_1p_{u_1} - u_2p_{u_2})}{u_1 - u_2} ,
\\
B_2=
\frac{\big(d\big(2u_1^2 + u_1u_2 - u_2^2\big) - 2cu_2u_1^2 + 2bu_1^2u_2^2 + u_1u_2^3(u_2 - 3u_1)\big)u_1\varphi(u_1)p_{u_1}^2}{(u_1 - u_2)^2}\\
\hphantom{B_2=}{}
+\frac{2u_1u_2(u_1 + u_2)\varphi(u_1)\varphi(u_2)p_{u_1}p_{u_2}}{(u_1 - u_2)^2}\\
\hphantom{B_2=}{} +\frac{\big(d\big(2u_2^2+ u_1u_2 -u_1^2\big) - 2cu_1u_2^2 + 2bu_1^2u_2^2 + u_2u_1^3(u_1 - 3u_2)\big)u_2\varphi{u_2}p_{u_2}^2}{(u_1 - u_2)^2} .
\end{gather*}
According \cite{bm,dor,win00} this Hamiltonian defines magnetic flow on the sphere.

\section{Conclusion}
We discuss a relatively simple map $\rho_b$ (\ref{g-trans2}) preserving the form of Hamiltonian
\[
\mathcal H=b_1p_{x_1}^2+\dots +b_np_{x_n}^2 ,\qquad b_i\in\mathbb R ,
\]
and the Dirac--Poisson bracket (\ref{d-br}) on cotangent bundle $T^*\mathbb S^{(n-1)}$ to the sphere.

Applying this map to the following Hamiltonian, which in terms of elliptic coordinates (\ref{ell-sph}) has the form
\[
T=\mathcal S_1+\dots+\mathcal S_{n-1} ,\qquad \mathcal S_k=u_k^m\prod_{i=1}^n (u_k-a_i)p_{u_k}^2 ,\qquad m=0,1,\dots ,
\]
we obtain polynomials of the second order in momenta at $b_i=a_i^\ell$, $\ell=0,1,\dots$,
\[\rho_b(T)=\sum_{i,j=1}^n \mathrm g^{ij}_b(x) p_{x_i} p_{x_j} .\]
Because these polynomials commute with $n$ independent, non-polynomial functions $\rho_b(\mathcal S_k)$, they determine a set of equivalent metrics $\mathrm g_b(x)$ on the sphere. By adding various potentials $V_b$ to these equivalent geodesic Hamiltonians $\rho_b(T)$ we can construct different integrable flows and different metrics (\ref{mau-metr}) on the sphere.

 The main problem is how to get a set of functions on $\rho_b(\mathcal S_k)$ which are polynomials in momenta. In this note, we study two-dimensional sphere when $n=3$ and prove that second-order polynomial $\rho_b(\mathcal S_1+\mathcal S_2)$ commutes with a polynomial of fourth order in momenta $\rho_b(\mathcal S_1\mathcal S_2)$. In further publications, we will present a similar result for equivalent metrics on the three-dimensional 3D sphere.

 Another interesting problem is to consider canonical transformations preserving Hamiltonian of the form
 \[
\mathcal H=b_1p_{x_1}^2+\dots+ b_np_{x_n}^2+V(x) ,\qquad b_i\in\mathbb R ,
\]
which were obtained for different partial cases in \cite{ts15,ts15a,ts17ell,ts17v, ts17c}.

\appendix

\section{The Maupertuis principle}
 In modern invariant, coordinate-free Hamiltonian mechanics \cite{arn,kup77}, an integrable system is defined as a Lagrangian submanifold in which $n$ parameters are considered as functions on $2n$-dimensional symplectic manifold. In a generic case, the Lagrangian submanifold depends on $m>n$ parameters and gives rise to a family of $C^n_m$ integrable systems with common trajectories.

 In traditional Hamiltonian mechanics, there are several coordinate-dependent descriptions of the integrable system with common trajectories \cite{bl12}, and the Maupertuis principle is the oldest of them. Roughly speaking, the Maupertuis or Jacobi--Maupertuis principle says that trajectories of the natural Hamiltonian systems are geodesics for the suitable metrics on configuration space, see \cite{bf05,bol95,ts99, ts01} and references within.

Below we present known technical construction of the geodesic Hamiltonians in a suitable to our purpose form.
Let us take the Hamilton function in the so-called natural form
\[
H=T+V(q) ,\qquad T=\sum_{i,j} \mathrm g_{ij}(q) p_ip_j ,\]
where potential $V(q)$ is a function on coordinates $q$ and $c$. Suppose that $H$ commutes with a~sum of the homogeneous polynomials of $m$-order in momenta
\[K=\sum_{m=0}^N K_m,\]
where $N$ is an arbitrary integer number, all terms in the polynomial $K$ have the same parity.

From $\{H,K\}=0$ follows that geodesic Hamiltonian
\[\tilde{T}=\sum_{i,j}\tilde{g}_{ij}(q) p_ip_j=\dfrac{T}{h-V} ,\qquad \tilde{\mathrm g}(q)=\frac{\mathrm g(q)}{h-V},\]
where $h$ is a constant, commutes $\big\{\tilde{T},\tilde{K}\big\}=0$ with a sum of the homogeneous polynomials of $m$-order in momenta
\[
\tilde{K}=K_m+\tilde{T}K_{m-2}+\tilde{T}^2K_{m-4}+\cdots .
\]
Indeed, we can rewrite equation $\{H,K\}=0$ as a set of equations
\[
\{ T,K_j \}+\{V,K_{j+2}\}=0 ,\qquad j=m,m-2,\dots ,\qquad K_{m+2}=K_{-1}=K_{-2}=0
\]
by using Euler's homogeneous function theorem.
Substituting these equations into
\begin{align*}
\big\{\tilde{T},\tilde{K}\big\}&= \big\{\tilde{T},K_m\big\}+\tilde{T}\big\{\tilde{T},K_{m-2}\big\}+\tilde{T}^2\big\{\tilde{T},K_{m-4}\big\}+\cdots \\
&= \dfrac{\{T,K_m\}}{h-V}+\left(\dfrac{T}{(h-V)^2}\{V,K_m\}+\dfrac{\tilde{T}}{h-V}\{T,K_{m-2}\}\right)+\cdots \\
&= 0 +\dfrac{T}{(h-V)^2}\big(\{V,K_m\}+\{T,K_{m-2}\}\big)+\cdots=0 ,
\end{align*}
and grouping terms of the same order in momenta we directly verify that $\tilde{T}$ commutes with $\tilde{K}$.

\subsection*{Acknowledgements}
We are very grateful to the referees for thorough analysis of the manuscript, constructive suggestions and proposed corrections, which certainly lead to a more profound discussion of the results.
The work was supported by the Russian Science Foundation (project 21-11-00141).

\pdfbookmark[1]{References}{ref}
\LastPageEnding

\end{document}